\documentstyle{lamuphys}
\makeatletter
\let\chapter\hid@chapter
\makeatother
\input{epsf.sty}

\begin{document}
\pagenumbering{arabic}
\title{Chemical evolution of low mass disc galaxies}

\author{Mercedes\,Moll\'{a} and Jean-Ren\'{e}\,Roy}

\institute{D\'{e}partement de Physique and Observatoire du mont M\'egantic,
Universit\'{e} Laval, Qu\'{e}bec, 
G1K 7P4 Qc, Canada}

\maketitle

\begin{abstract}

We show that the multiphase chemical evolution model reproduces the
correlations obtained along the spiral sequence, 
dwarf galaxies included.
However the apparent spatial chemical uniformity observed in
 some irregular galaxies cannot be reproduced with it.
An evolutionary model has been developed and tested
to explain flat gradients.  Preliminary results, obtained with a
new code including supernova winds and radial flows,
suggest that radial flows are probably responsible for
this  uniformity.

\end{abstract}
\section{The multiphase model: summary and generic results}

The multiphase model has already been used for the solar neighborhood
(Ferrini et al. 1992), for both the Galactic bulge and disc (Ferrini et
al. 1994; Moll\'{a} \& Ferrini, 1995) and for some nearby
spiral galaxies (Moll\'{a}, Ferrini, \& D\'{\i}az, 1996,
hereinafter MFD96).
The model starts with a sphere of primordial gas whose total mass
$M_{tot}$ is calculated using rotation curves. This mass
collapses onto an equatorial plane on a timescale $\tau_{0}$. The
sphere is divided into concentric cylindrical regions, each of which
having a halo and a disc zone.
The star formation rate in the disc is considered a two step
process: First, the diffuse gas forms molecular clouds at a rate which
depends on the efficiency $\epsilon_{\mu}$.  Then stars form from
cloud-cloud collisions with efficiency
$\epsilon_{h}$, or by the interactions of
massive stars with molecular gas clouds with efficiency
$\epsilon_{a}$.

When applied to different spiral galaxies (MFD96), the
characteristic values of $\tau_{0}$ and efficiencies $\epsilon_{h}$ and
$\epsilon_{\mu}$ change, depending on the total mass of the galaxy,
the Arm Class and the Hubble type respectively. 
Radial distributions of oxygen abundance, atomic and molecular gas
surface densities and star formation rates, which are used as
constraints for the models, are reproduced with larger $\tau_{0}$ values
for less massive galaxies and efficiencies $\epsilon_{\mu}$, and
smaller $\epsilon_{h}$ values for later type galaxies.  Thus
star formation histories differ from galaxy to galaxy and from region 
to region in a given galaxy. Generic trends are consistent with observed
correlations for large spirals (Vila-Costas \& Edmunds, 1992;
Zaristky, Kennicutt \& Huchra 1994; Oey \& Kennicutt 1994).

\section{The model for the low-mass end of the spiral sequence}

Recently obtained data on low-mass and dwarf galaxies
show that these trends are maintained: low-mass
irregular galaxies and large spirals fall on the same
sequence (Hoffmann et al. 1996
-- HOF96, Broeils \& Mc Rhee 1997 -- BR97,
McGaugh \& de Block 1997 --  GB97).
 
Wishing to see if the multiphase model reproduce these correlations, 
we simulated
dwarf galaxies as unevolved systems: low in luminosity and irregular in
their optical appearance, a star formation rate which does not follow
the spiral wave but with recent bursts, low metallicities, blue
colors and high gas fractions.  We chose smaller values of $\epsilon_{\mu}$ and
$\epsilon_{h}$  than those used for NGC~598 and
NGC~300, the latest type galaxies modeled by MFD96.  For low-mass
galaxies, $\tau_{0}$ is larger due to the smaller total mass.

\begin{figure}
\vspace{5cm}
\includegraphics{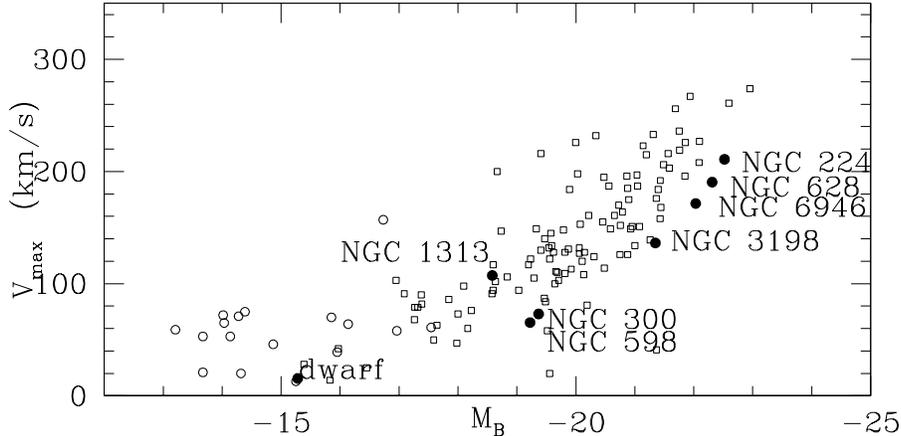}
\caption{ The maximum rotation velocity  V$_{max}$  {\sl vs.} M$_{B}$.
Open symbols  are the data taken from HOF96 and from BR97
 Filled circles are the multiphase  model results.}
\end{figure}

Using these  models, we investigated the behavior along the whole
spiral sequence.  Dynamical masses
have larger values for outer disc regions.  The ratio
$\rm M_{dark}/M_{lum}$ is higher for low-mass galaxies than for
large spirals, and it increases with  radius. These effects are reflected
by the multiphase model in leading to longer
 $\tau_{0}$ when the total mass is lower;
this timescale is assumed to be increasing with
radius. The correlation observed between the maximum
rotation velocity $V_{max}$ and the total magnitude $M_{B}$ is
reproduced by the multiphase models (Figure 1).  The model results
represent a range of galaxy types, from the earlier type
 galaxy NGC~224 ($T=3$) 
to the intermediate type galaxies NGC~628 and
NGC~6946 ($T=5, 6$), the later type galaxies, NGC~598 and
NGC~300 ($T=6, 7$) and finally the magellanic irregular galaxy
NGC~1313 ($T \geq 8$).

Other correlations refer to gas quantities and  total magnitudes
or luminosities. The relation between the gas fraction
$f_{B}$ and $M_{B}$ is shown in Figure 2a. While being less massive,
galaxies with lower
luminosities have lower absolute gas mass, but have the largest gas 
mass fraction. 
Spiral galaxies with lower luminosities are also those with later
morphological types. It results in a correlation between 
$f_{B}$ and the Hubble type T.

The characteristic oxygen abundance 12 + $log$
(O/H) is related to $M_{B}$ for spiral galaxies: the more
luminous ones have higher abundances (Skillman, Kennicutt \& Hodge 1989
 -- SZH89; Zaritsky, Kennicutt \& Huchra 1994 -- ZKH94;
Ritcher \& McCall 1995 -- RMC95; R\"{o}nnbavk \& Bergvall 1995 -- RB95;
Garnett et al. 1997 -- GSSSD97). The same trend is valid for the 
luminosity range
$M_{B}= -10$ to $M_{B}= -20$. Models reproduce this (Figure 2b).

\begin{figure}
\vspace{5cm}
\includegraphics{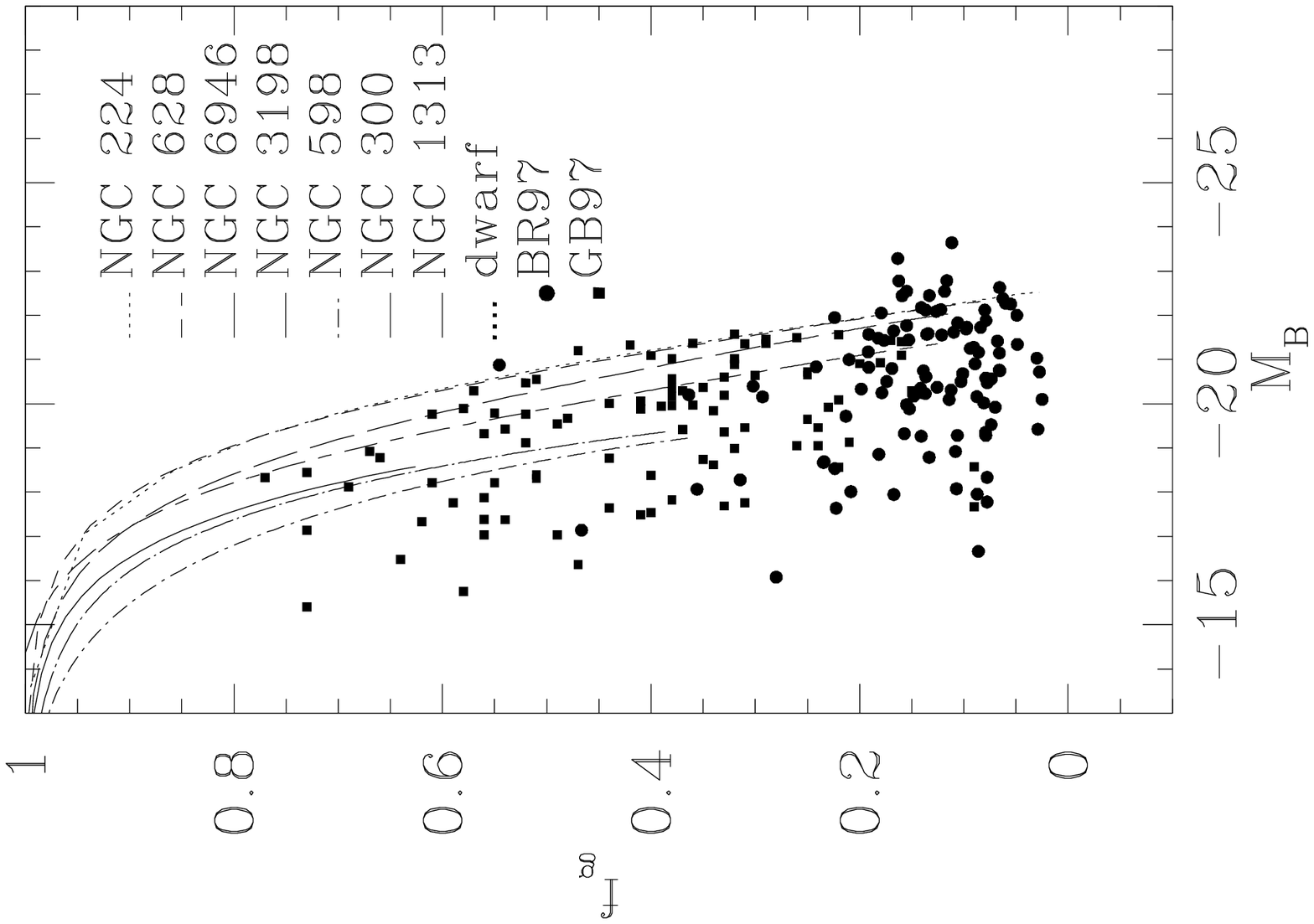}
\includegraphics{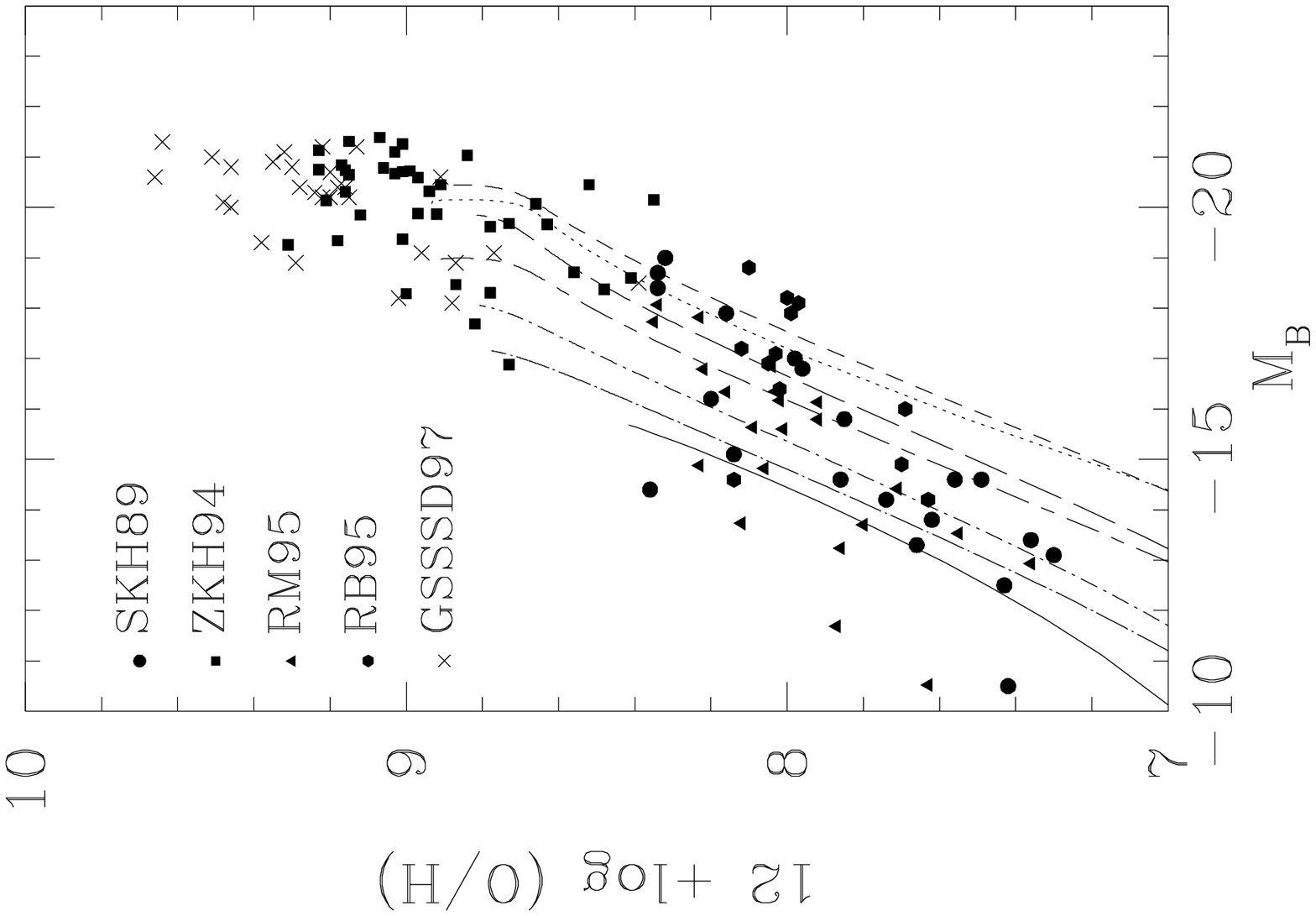}
\caption{a) The gas fraction $f_{B}$ {\sl vs.} $M_{B}$ for spiral galaxies. 
Symbols are the data taken from BR97 and from GB98. b) The relation of
[O/H] {\sl vs.} M$_{B}$. Data are from SKH89, ZKH94, RMC95,
RB95 and GSSSD97. Lines correspond to our models for each type of galaxies.}
\end{figure}

There are two other known facts: 1) early type galaxies
have higher abundances in their discs than later type ones; 2)
radial gradients are steeper in late-type galaxies than in
early-type ones. Both results are
reproduced by models in Figure 3: Fig. 3a shows the characteristic
O/H abundances (open symbols) versus T; in Fig. 3b,
radial gradients are plotted against T.  
A trend of steeper radial gradients with less 
evolved and later type galaxies is obvious.

Therefore, most of the characteristics of low-mass irregular and normal
large spirals are reproduced by the multiphase model based on
rather simple physical assumptions. For example,
the total mass of a galaxy
defines the rate of collapse and disc formation. The star
formation rate depends on the gas mass fraction in the disc, and also
on the arm class or morphological type.  The more massive galaxies evolve
more rapidly and reach higher abundances quickly. The less massive
 galaxies take longer to form the disc and some of them are just now 
reaching their peak star
formation in their centers; this leads to a steep radial gradient of
abundances due to the production of oxygen in this region.  It is also
possible to have two galaxies with similar total masses but different
types of spiral density wave, allowing different star formation histories. 
This could explain the large dispersion of the data in most of the
correlations.

\begin{figure}
\vspace{5cm}
\includegraphics{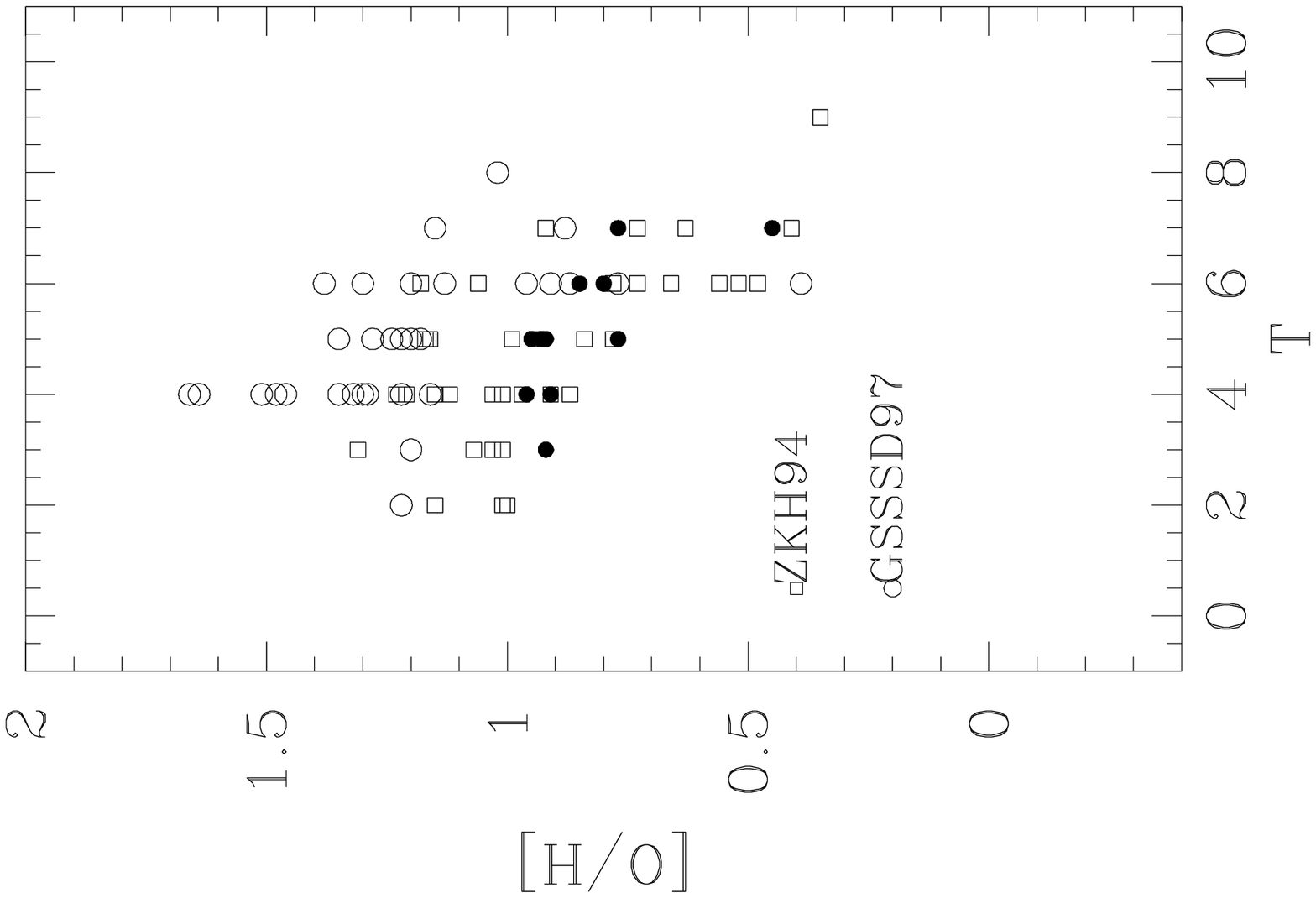}
\includegraphics{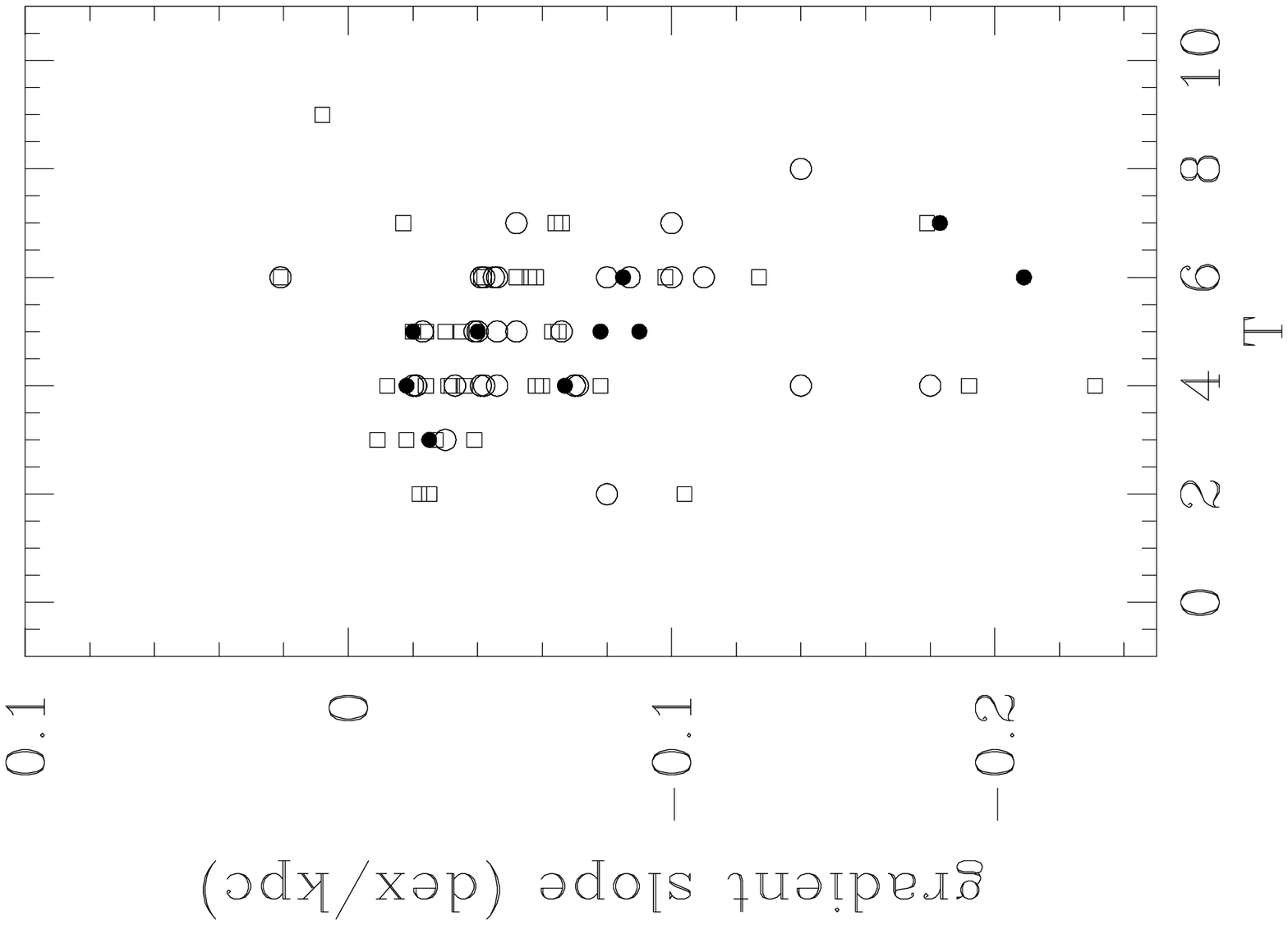}
\caption{a). Oxygen abundance [O/H] vs. Hubble type T.
b). Radial gradient for [O/H].
Open symbols correspond to data from ZKH94 and from GSSSD97.  
Filled circles are the multiphase model results.}
\end{figure}

\section{The abundance uniformity in dwarf galaxies.} 

However, radial gradients disappear when spiral
structures no longer exists (Edmunds \& Roy, 1993). While in normal
galaxies radial gradients steepen for later type galaxies, the low
mass irregular galaxies show a large uniformity of abundances over their
discs. This is difficult to understand especially because a burst 
of star formation is generally occuring in
their centers (Skillman, Dohm-Palmer \& Kobulnicky 1998 and references 
therein).

For example, the galaxy NGC~1313, considered as a galaxy in
transition between a magellanic type  and  a normal spiral such
as the Sc galaxies NGC~598 (M~33) or NGC~300, and similar to both in
total mass, shows a flat radial gradient of oxygen abundance (Walsh \&
Roy 1997).  By choosing efficiencies smaller than those
chosen for NGC~598 and NGC~300 and a similar collapse timescale, 
the multiphase model (see Moll\'{a} \& Roy 1999 for details)
produces a steep radial gradient contrary to what is observed. 
If we assume a constant $\epsilon_{\mu}(R)$ to take into account
that it is an irregular galaxy 
without a strong spiral wave system, a flat radial gradient is obtained but
the theoretical star formation radial profile $\Psi (R)$ is
much flatter than observed.
The only way to reproduce the observed $\Psi (R)$ is to have a
burst of star formation in the center of the galaxy. But then, [O/H]
is higher than observed and the radial gradient steepens.

This is a well known problem and some solutions have been 
suggested (see Kobulnicky \& Skillman 1996) to explain why the
heavy elements may not be observed in regions with young massive
stars. The global star formation conspiracy is not valid with a
maximum for $\Psi (R)$ in the center. 
We have developed a new version of the
 multiphase chemical evolution model which allows
mass loss by supernova (SN) explosion winds and the
possibility of mass exchanges between regions. As a first test,
we simulated selective  mass loss: the oxygen goes to a hot phase
and remains there for a time (the hidden ejecta hypothesis),  
or it is expelled into the halo at large galactocentric distances 
to fall back later onto other disc regions following  Charlton \& 
Salpeter (1996). Preliminary results 
indicate that this scenario does not work: the
star formation increases in the center (without increasing  [O/H])
but only for early epochs.  Morever, the $\Psi(R)$ profile has the same
shape than before and the gas density ends up lower than observed.

Until now, the multiphase model has not included the
effect of radial flows induced by central bars.
Barred galaxies have flatter radial abundance gradients than normal 
spirals (see Roy 1996) and there is a direct relation between the bar 
strength and the the radial gradient amplitude.  When a bar
appears, radial flows of gas are produced by the
non-axisymmetric gravitational potential. These flows
mix the interstellar medium and flatten the radial gradient
over a timescale of ~1 Gyr (Friedli, Benz \& Kennicutt 1994; Edmunds
\& Greenhow, 1995). Therefore, irregular and magellanic type
galaxies may have or have had a bar. For
NGC~1313, this possibility is appropriate to explain the flat
gradient and the star formation with a maximum in the center: the
gradient can become flat, assuming a radial flow of gas with 
a mean effective velocity of
$\sim 20$ km/s lasting a few Gyr (2-3 Gyr) and having 
started about 5 Gyr ago, when the galaxy was 8 Gyr old.
We are now exploring this scenario with a new model which
allows the exchange of matter between radial zones.

%
%

\end{document}